\documentclass[12pt]{iopart}
\usepackage{graphicx}
\usepackage{color}
\newcommand{\red}{\color{black}}
\newcommand{\blk}{\color{black}}

\newcommand{\beq}{\begin{equation}}
\newcommand{\eeq}{\end{equation}}
\newcommand{\bqa}{\begin{eqnarray}}
\newcommand{\eqa}{\end{eqnarray}}

\newcommand{\bra}[1]{\langle{#1}|}
\newcommand{\ket}[1]{|{#1}\rangle}
\newcommand{\ip}[2]{\langle{#1}|{#2}\rangle}

\newcommand{\hei}{Heisenberg}

\newcommand{\singlecol}{\end{multicols}
     \vspace{-0.5cm}\noindent\rule{0.5\textwidth}{0.4pt}\rule{0.4pt}
     {\baselineskip}\widetext }
\newcommand{\doublecol}{\noindent\hspace{0.5\textwidth}
     \rule{0.4pt}{\baselineskip}\rule[\baselineskip]
{0.5\textwidth}{0.4pt}\vspace{-0.5cm}\begin{multicols}{2}\noindent}

\newcommand{\ea}{{\em et al.}}

\begin{document}

\title{A double-slit `which-way' experiment on the
complementarity--uncertainty debate}
\author{R.~Mir$^{1}$, J.~S.~Lundeen$^{1}$, M.~W.~Mitchell$^{2}$, A.~M.~Steinberg%
$^{1}$,  J.~L.~Garretson$^{3}$, H.~M.~Wiseman$^{3}$}
\address{$^{1}$Centre for Quantum Information \& Quantum Control and Institute for
Optical Sciences, Department of Physics, 60 St.~George Street, University of
Toronto, Toronto ON M5S\\
$^{2}$ICFO - Institut de Ci\`{e}ncies Fot\`{o}niques, Jordi Girona 29, Nexus
II, 08034 Barcelona, Spain 1A7, Canada\\
$^{3}$Centre for Quantum Dynamics, Griffith University,
Brisbane, Queensland 4111 Australia. }

\begin{abstract}
A which-way measurement in Young's double-slit will destroy the interference
pattern. Bohr claimed this complementarity between
wave- and particle-behaviour is enforced by Heisenberg's uncertainty
principle: distinguishing two positions a distance $s$ apart transfers a
random momentum $q\sim \hbar /s$ to the particle.  This claim has 
been subject to debate: Scully \ea\ asserted that in some situations 
interference can be destroyed with no momentum transfer, while 
Storey \ea\ asserted that Bohr's stance is always valid. We address this 
issue using the experimental technique of weak measurement. We measure a 
distribution for $q$ that spreads well beyond $[-\hbar/s,\hbar/s]$, but nevertheless has a 
 variance consistent with zero. This weak-valued momentum-transfer 
 distribution $P_{\rm wv}(q)$ thus reflects both sides of the debate. 
\end{abstract}

\maketitle
 
 \section{Introduction}

An interference pattern forms when it is impossible to tell through which of
two slits a quantum particle travelled to a distant screen. Conversely, performing a
which-way measurement (WWM) to determine which of these two paths the particle took
destroys this pattern.  This choice of exhibiting wave-like or 
particle-like behaviour was called complementarity by Bohr \cite{Boh28}.  
 
 In his debates with
Einstein, Bohr \cite{BohrEinst} argued that complementarity was enforced the 
 (then newly discovered)  
Heisenberg uncertainty principle. By this he meant 
 the measurement--disturbance relation which Heisenberg formulated in 
 1927: in a measurement of position, Planck's constant $h$ gives a lower bound 
on  the product of the ``precision with which the position is known'' and 
 ``the discontinuous change of momentum'' \cite{Hei27}.
 In the context of the double-slit experiment, Bohr argued that  
a measurement able to distinguish two positions a distance $s$ (the slit separation)
 apart must produce an  ``uncontrollable change in the momentum''  
 $q \sim h /s$. This is just the magnitude required to wash out the interference fringes, 
 which have a period of $h/s$ in momentum space.
  
 Bohr's argument was famously reiterated by Feynman~\cite{FeyLeiSan65}, who said 
 ``No one has ever thought of a way around the uncertainty principle." 
However in 1991, Scully, Englert, and Walther \cite{ScuEngWal91} proposed a specific WWM
that, according to their calculations, transfers essentially no momentum. This seemed to
 show that complementarity is more fundamental than the uncertainty principle. 
 Their calculation consisted of a proof that a {\em
single-slit} wavefunction was essentially unchanged by their WWM. 
 
  The argument of Scully \ea\ was not accepted by Story, Collett, Tan and Walls \cite{StoTanColWal94}. They 
proved a general theorem showing, they claimed, that any WWM causes a 
momentum transfer at least of order $\hbar /s$, so that the uncertainty 
 principle is indeed relevant to double-slit experiments. They identified 
 the momentum disturbance as occurring in the convolution of the 
 momentum probability {\em amplitude} distribution. Observationally, 
 their theorem means that if the initial state were a {\em momentum eigenstate} 
 then the final (i.e.~after the WWM) momentum distribution 
 would have a width \cite{fn1} of at least $\hbar/s$ \cite{Wis97a}. 

In this paper we present the first experimental work to address this debate 
\cite{ScuEngWal91,StoTanColWal94,Nature95} 
about momentum disturbance by a WWM in a double-slit apparatus \cite{NoDbslt}.
We use a WWM akin to that proposed by Scully \emph{et
al.}, but using photons rather than atoms. Using a technique proposed recently 
by one us \cite{Wis03}, we measure a {\em weak-valued} probability
distribution for $q$. Our measured distribution for 
 $q$ has a width \cite{fn1} clearly greater than $\hbar/s$, 
 but has a variance consistent with zero, thus exhibiting   
 features characteristic of both sides in the debate. 
 \red This is possible only because a weak-valued probability can take 
negative values \cite{Wis03}.\blk 
 
 This paper is organized as follows. In Sec.~2 we discuss the differing  
 concepts of momentum transfer that have been used in the debate,
 including the weak-valued momentum-transfer distribution. In Sec.~3
 we explain this last concept in detail, using only concepts understandable 
 to a classical physicist. It is on this basis that we say that we have {\em directly
 observed} the momentum-transfer distribution in our experiment, described in Sec.~4. 

\section{Concepts of Momentum Transfer}
 
That both sides in the debate \cite{ScuEngWal91,StoTanColWal94,Nature95} had
valid claims was first pointed out in Ref.~\cite{WisHar95}. The disagreement
came from the fact that the two groups were using different concepts of
momentum transfer. One might have thought that momentum transfer or disturbance 
 was defined by Heisenberg in 1927 \cite{Hei27}, and so should have no ambiguity. 
In fact, \hei's measurement--disturbance relation, as appealed
 to by Bohr and Feynman, used no quantitative definition of momentum disturbance 
 at all. This is in contrast to the other uncertainty 
relation formulated in Ref.~\cite{Hei27}, referring to ``simultaneous determination of two 
canonically conjugate quantities''. This relation between simultaneously determined uncertainties   was  immediately put on a rigorous footing by Weyl \cite{Wey28}, using standard deviations in the familiar relation 
 $\sigma(x)\sigma(p) \geq \hbar /2$. As \hei\ recognized in 1930 \cite{Hei30}, 
 it is only in the case that the particle is initially in a momentum eigenstate that the 
 simultaneously-determined-uncertainty relation can be used to derive a rigorous measurement--disturbance relation --- see Ref.~\cite{Wis98c} for a discussion. Thus, in the context of the double-slit apparatus, one cannot use the simultaneously-determined-uncertainty relation
as a basis for a measurement--disturbance relation. 
 
The definitions of momentum disturbance 
adopted by Scully \ea\ and by Storey \ea\ were both reasonable. 
Moreover, both concepts agreed for the cases of \emph{classical} momentum 
transfers \cite{WisHar95}. By this phrase, Wiseman and Harrison meant a momentum 
transfer that could be described as a random momentum kick, drawn from some 
(positive) distribution $P_{\rm cl}(q)$ of momenta $q$. Examples of WWMs 
resulting in classical momentum transfer include all those 
discussed by Bohr \cite{BohrEinst} and Feynman \cite{FeyLeiSan65}. 
For WWMs with classical momentum transfer, the final momentum probability 
distribution is obtained by convolving the initial  momentum {\em probability distribution} 
with $P_{\rm cl}(q)$ \footnote{This is as opposed to a convolution of the momentum {\em probability amplitude distribution} in the general case as analysed by Storey \ea\ \cite{StoTanColWal94}.}. As a result, the momentum transfer is independent 
of the initial state and can be quantified by the increase in the variance of the particle's 
momentum, which will equal the variance of $P_{\rm cl}(q)$ \cite{Wis97a}.

The WWM of Scully \emph{et al.} does {\em not} result in a classical momentum transfer. 
This is what allows their result, that a 
\emph{single-slit wavefunction} would be unchanged by their WWM 
--- in particular, it suffers no increase in momentum variance\footnote{In fact, 
measurements of this kind also cause no change in the variance (or indeed in any 
of the moments) of the momentum probability distribution of the {\em double-slit} wavefunction \cite{Wis97a},
despite the fact that the momentum probability distribution itself is changed drastically
due to the disappearance of the fringes. The same effect (invariance of the momentum moments) 
occurs in the Aharonov-Bohm effect, despite the fringe shift, as shown by Aharonov and 
co-workers~\cite{AhaPenPet69}; see also the discussion in Ref.~\cite{Wis97a}. 
[Ref.~\cite{AhaPenPet69} shows that there {\em is} however a change in the 
{\em modular momentum}.] 
The relation of the WWM of Scully \ea\ to the Aharonov-Bohm 
effect emphasizes its nonclassical nature \cite{AhaPenPet69,Wis97a}.
It should be noted, however, that the invariance of the momentum moments is not 
readily accessible experimentally because for slits with sharp edges (as in our apparatus), 
the variance of the initial and final momentum probability distributions is undefined  
{\em even with} a regularization procedure (as discussed in section 3 in the context of 
the variance of the weak-valued momentum-transfer distribution) \cite{Gar04}.}.
But at the same time, the WWM of Scully \ea\ does not evade the theorem of Storey \emph{et al.} involving momentum-transfer probability amplitudes. This theorem  
implies that any WWM would disturb a \emph{momentum eigenstate}, resulting 
in a final momentum probability distribution with a width \cite{fn1} of at least $\hbar/s$ \cite{Wis97a}. 


Although the calculations of Scully \ea\ and Storey \ea\ are not in conflict, it is 
unsatisfying that their physical predictions
require experiments (with a single-slit wavefunction and momentum eigenstate
respectively) that are incompatible with each other and with the \emph{%
double-slit} experiment that they are supposed to illuminate. In contrast,
the weak measurement technique that we  outline in the next section allows us to observe 
directly the momentum transfer while carrying out the original
double-slit experiment \cite{Wis03}. Moreover, this weak measurement
technique is unique in allowing aspects from the calculations from both sides of the debates
to be seen in a single momentum-transfer distribution \cite{Wis03}.

 \section{Theory}
 
Consider a double-slit experiment in which the slits are separated 
 in the horizontal ($x$) direction, with a WWM following the slits. 
 We are interested in the change in the particle's momentum from its initial 
 state (just after the slits) to its final state (after the WWM). 
 For a physicist ignorant of quantum mechanics, an obvious  way to probe this momentum 
 transfer would be to `tag' particles with an initial momentum $p_{i}$ 
  using a parameter of the particle uninvolved in the interference
effect. For example, one could   
tag particles by inducing in them a vertical displacement $D$.
  (This is related to the technique we use in the experiment, as described below.) 
  Then, after the WWM, these particles would be detected at the screen with final momentum 
$p_{f}$. The difference, $q=p_{f}-p_{i}$, would be the momentum transfer.
  
  One can arrive at a probability distribution for $q$ by selecting the subset of particles with
final momentum $p_{f}$ and counting the number of these that are tagged. In
this post-selected subset, $(\textrm{\# tagged})/(\textrm{total \#})={\rm Pr}(p_{i}|p_{f})
$, the probability that a particle began with $p_{i}$ given that it was
later found with momentum $p_{f}$.  One repeats this for every combination
of $p_{i}$ and $p_{f}$ to attain the unconditional joint probability
distribution $P(p_{i},p_{f})=P(p_{i}|p_{f}) P(p_{f}).$ From this one
finds the probability for a momentum transfer $q$, averaged over the initial
(or final) momentum of the particle:
\begin{equation}
P(q)\equiv \sum_{p_{i}}\left[ P(p_{i},p_f)\right]_{p_f=p_{i}+q} = \sum_{p_i} \left[P(p_{i}|p_{f}) P(p_{f})\right]_{p_f=p_{i}+q}. \label{prob sum}
\end{equation}%
 Here we are treating $p_{i}$ as discrete, as is appropriate for our
experimental apparatus, described later. That is, $P(p_i|p_f)$ is  
in fact a probability, whereas strictly $P(q)$ is a probability {\em density}. 

For classical particles the procedure just described is completely
equivalent to the following. Instead of counting tagged particles, one
measures the \emph{average vertical displacement} of the whole subset, $d=[(\textrm{\# tagged})
 \cdot D+(\textrm{\# untagged})\cdot 0]/(\textrm{total \#})$, so that $%
P(p_{i}|p_{f})=d/D$. In this experiment, we use this variation because it
does not require us to know whether a particular detected particle was
tagged (i.e.~had momentum $p_{i}$) or not. Classically, determining the
momentum of a particular particle is harmless, but in quantum mechanics it
would collapse the state to a \emph{momentum eigenstate} $\left| {p}%
_{i}\right\rangle$. 
In effect, the tagging procedure would be a `strong' measurement of initial
momentum and the ensuing collapse would disturb the very process we wish to
investigate, the effect of the WWM on the double-slit wavefunction. Thus,
for the quantum experiment, we need a way of  reducing this
disturbance to an arbitrarily low level. 
 
 The solution is to make a `weak' measurement, reducing our ability to
discriminate whether a particular particle had momentum $p_{i}$ or not. We
do this by making the induced displacement $D$ small compared to the
vertical width $\sigma $ of the particle's wavefunction. A classical
physicist would regard this as merely reducing the signal-to-noise ratio of the
measurement, and would still interpret the result $d/D$ as giving 
 $P(p_{i}|p_{f})$. The reduced signal-to-noise can be overcome simply by 
running more trials.

The quantity $d/D$ is the average value of a weakly measured quantity
post-selected on a particular outcome. This is what is known as a \emph{weak
value} \cite{AhaAlbVai88}. It can be shown theoretically that a general weak
value (in the limit $D/\sigma\rightarrow 0$) can be calculated by: 
\begin{equation}
\hspace{-5ex}{\phantom{(X_{w})}}_{\phi }\!\left\langle {X_{\mathrm{w}}}%
\right\rangle _{\psi }=\mathrm{Re}\frac{\langle {\phi }|\hat{U}\hat{X}|{\psi 
}\rangle }{\langle {\phi }|\hat{U}|{\psi }\rangle }.  \label{weakvalgen}
\end{equation}%
Here the initial state of the system is $\left| {\psi }\right\rangle $,  the postelected state 
 is $\left| {\phi }\right\rangle $, and $\hat{X}$ is the weakly
measured observable.In the above procedure, these are the double-slit
wavefunction, the final momentum state $\left| {p}_{{f}}\right\rangle $, and
the projector for the discretized initial momentum $\left| {p}%
_{i}\right\rangle \left\langle {p}_{i}\right|$, respectively.Evolution
after the weak measurement is given by $\hat{U}$, typically unitary, but in
our case, an operation describing the measurement of the particle by the WWM
device~\cite{Wis03,Gar04}.The generality of weak values has made them
useful tools to analyze a great variety of quantum phenomena \cite%
{RitStoHul91,Ste95,Aha02,Mol01,Wis02a,RohAha02,Brun03,Sol04,Res04,Pry05,Wis07}.

If no post-selection is performed then it can be shown that the average of
the weak measurement result is the same as for a strong measurement, $%
\langle {\psi }|\hat{X}|{\psi }\rangle$. For $\hat{X}$ equal to the
projector $\ket{p_i}\bra{p_i}$, this expectation value is equal to initial
momentum probability, $P(p_i) = \ip{\psi}{p_i}\ip{p_i}{\psi}$. 
 However, in the case of post-selection on
state $\left| {\phi }\right\rangle $, the weak value may lie outside the
eigenvalues of $\hat{X}$ \cite{AhaAlbVai88}, a prediction that was quickly
verified experimentally~\cite{RitStoHul91}. In particular, if we weakly
measure a projector, the weak value can lie outside the range [0,1].
 This is of course impossible for a true probability.  To make this distinction, we call
the weak value of a projector a \emph{weak-valued }probability (WVP).\ The
fact that a WVP can be negative enables it to describe states and processes
which \emph{require} a quantum description, similar to other
quasi-probabilities such as the Wigner function \cite{Wis97a}. 

The application of WVPs to momentum transfer in WWMs was first considered in
Ref.~\cite{Wis03}. Our quantity $d/D$, which a classical physicist would
call the conditional probability $P(p_{i}|p_{f})$, is, in the limit $%
D/\sigma \to 0$, exactly the conditional WVP: 
\beq
d/D \to P_{\rm wv}(p_i|p_f) = \hspace{-5ex}{\phantom{(X_{w})}}_{p_f}\!
\Big\langle {\ket{p_i}\bra{p_i}_{\mathrm{w}}}\Big\rangle _{\psi }
\eeq
We manipulate this result according to Eq.~(\ref{prob sum}) just as a
classical physicist would, but we refer to the resulting quantity as the \emph{%
weak-valued} momentum-transfer distribution: 
\beq \label{wv prob sum}
P_{\mathrm{wv}}(q) = \sum_{p_i} \left[ P_{\rm wv}(p_{i}|p_{f}) P(p_{f}) \right]_{p_f=p_{i}+q}
\eeq
 It is
in this manner that we directly observe a momentum-transfer distribution: 
It is derived via a simple prescription, with no reference to quantum
physics, from measurements a classical physicist would understand.

Like a standard probability distribution, $P_{\mathrm{wv}}(q)$ as defined here 
integrates to unity. Moreover, its mean and variance exactly reflect the
change in the mean and variance of the momentum distribution that occurs as
a result of the WWM \cite{Gar04}. For WWMs that produce a classical momentum
transfer, $P_{\mathrm{wv}}(q)=P_{\rm cl}(q)$ and so is positive. However, 
for {\em nonclassical} momentum transfers,   $P_{\mathrm{wv}}(q)$ may 
go negative. We emphasize that the existence of negative values of $P_{\rm wv}(q)$ is not 
a flaw in the theory. Rather it is a necessary feature in order for $P_{\rm wv}(q)$ 
to reflect both sides of the debate, in that this distribution must have a width \cite{fn1} 
greater than $\hbar/s$, even though its variance $\int P_{\rm wv}(q) q^2 dq$ 
can be arbitrarily small. 
  
 One subtlety in relating the theory to experiment is that if the slits have sharp edges (as they do in our apparatus) then  $\psi(x)$ is not continuous. As a consequence, a regularization procedure
 is required to make the variance integral $\int P_{\rm wv}(q) q^2 dq$ converge \cite{Gar04}. For example, one can multiply $P_{\rm wv}(q)$ by an apodizing function $e^{- |q|/\kappa}$, calculate the integral, and then let $\kappa \to \infty$  \cite{Gar04}. If instead one replaces  this smooth cut-off with a sharp cut-off at $q=\pm q_{\rm max}$, then 
  the calculated variance diverges as $q_{\rm max}\to\infty$, oscillating between positive and negative values. Experimentally the regularization is not achievable with current techniques, 
 as it requires $P_{\rm wv}(q)$ to be measured to great accuracy over a very large range. 
 However the signature of a zero variance can be seen by calculating the variance from the data in the range $[-q_{\rm max},q_{\rm max}]$, and observing an oscillation from a positive value 
 to a negative value as $q_{\rm max}$ is increased. These oscillations are what ensures that the regularized variance evaluates to zero.

 \section{Experiment}

The experiment we report is the first to address the question of momentum
transfer by WWMs in a double-slit apparatus \cite{NoDbslt}. The experimental
apparatus is shown in Fig.~1. Since photons are non-interacting particles it is 
unnecessary to send only one through the apparatus at a time.Instead, we
use a large ensemble simultaneously prepared with the same wavefunction, as 
 produced by a single-mode laser. It
follows that the transverse intensity distribution of the beam is proportional to the
probability distribution for each photon. Treating the photons as particles,
a classical physicist would analyze the experiment using 
trajectories \cite{Gol80}. In this model, the transverse motion of the
photon is that of a free non-relativistic particle of mass {$m=h/c\lambda $}.
 
 \begin{figure}
\hspace{-0.25\textwidth}\includegraphics[width=1.25\textwidth,angle=-90]{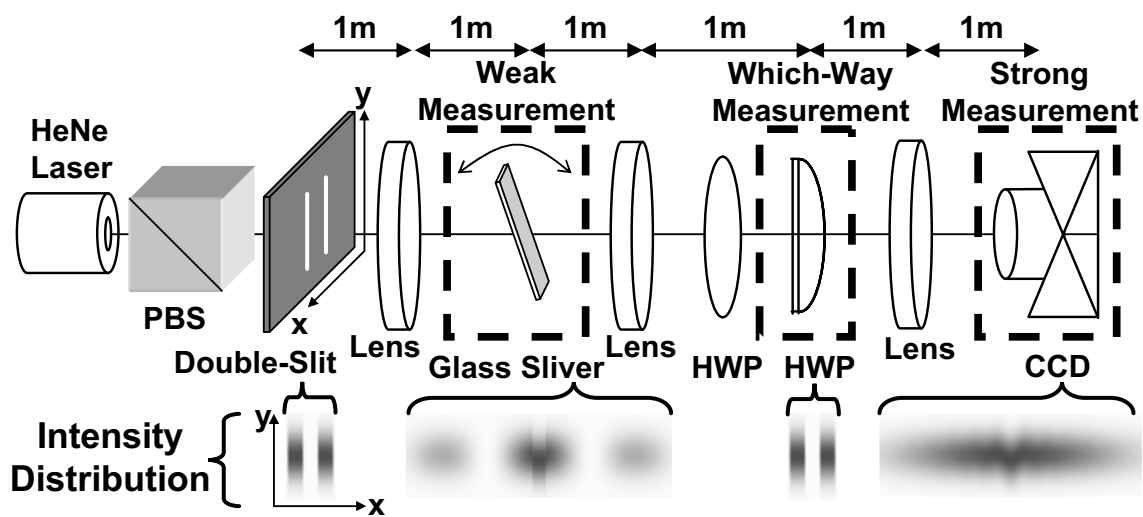}
\caption{Diagram of apparatus. The photons are prepared in the initial
state by a polarizing beam-splitter (PBS) and a double-slit aperture. 
They are then measured {\em three} times: the weak measurement 
via a $y$-displacement by the glass sliver at $p_i$; the which-way measurement 
via polarization rotation by the half-wave plate (HWP) at one slit; 
and the final strong measurement of $p_f$ by the 
 CCD camera. For details see text. Below the apparatus are shown calculated
 intensity distributions at the longitudinal positions indicated, with the $y$-displacement 
 exaggerated for clarity. }
\end{figure} 

The photon ensemble is produced by a 2 $\mathrm{mW}$ $\lambda =633\mathrm{nm}
$ HeNe laser that illuminates a double-slit aperture with a slit width of $%
w=40\mathrm{\mu m}$ and a center-to-center separation of $s=80\mathrm{\mu m}$. 
We call the long (vertical) axis of the slits $y$ and the axis joining
their centers $x$. We use $f=1\mathrm{m}$ focal-length lenses to switch
back and forth between position and momentum space for the photons. {These
can be treated as impulsive harmonic potentials} {in the classical particle
picture.} One metre after the first lens, the photon's $x$-position $x_{i}$
becomes equal to $(f/c)\cdot (p_{i}/m)$, where $p_{i}$ is its initial
momentum at the double-slit.Consequently, in the $x$-direction the
intensity distribution is that of the expected double-slit interference
pattern with a fringe spacing of $8.2\pm 0.1\mathrm{mm}$.\ In the $y$%
-direction, the intensity distribution is Gaussian with a $1/e^{2}$
half-width $\sigma =1.01\pm 0.01\mathrm{mm}.$

We tag the photons with a $y$-displacement $D$ $=$ $0.14\pm 0.01\mathrm{mm}$
($\ll \sigma $ ensures weakness) in a range of momenta $\Delta$ centered on $%
p_{i}.$ This displacement is induced by tilting an optically flat glass
sliver placed at $x_{i}$ with a width of $\delta =1.77\pm 0.02\mathrm{mm}$
in the $x$-direction and a thickness of $1.00\pm 0.25\mathrm{mm}$. That is,
the momentum resolution of our weak measurement of $p_{i}$ is $\Delta =
(m/cf)\cdot \delta \ll h/s$. If there is no momentum transfer we expect $P_{%
\mathrm{wv}}(p_{i}|p_{f})=1$ for $|p_{i}-p_{f}|<\Delta /2$ and $0$
otherwise. Any deviation from this represents a momentum disturbance.

To implement the WWM we must switch back to position space with a second $%
f=1 $m lens, in essence imaging the slits.Here, the photons pass through
a half-wave plate for fine alignment of their polarization.A second
half-wave plate in front of the image of just one of the slits flips the
polarization. That is, the photon polarization carries the WWM result,
destroying the double-slit interference.Since the spatial wavefunction is
unaltered, this is exactly the type of WWM Scully \emph{et al.} considered.

A third $f=1$m lens transforms back into momentum space, so that finally $%
x_{f}=(f/c)\cdot (p_{f}/m)$.\ Here we record the intensity distribution with
a movable CCD camera in an $x$-$y$ region of size $27.5{\mathrm{mm}}\times
2.70\mathrm{mm}.$ This was done for $x_{i}=n\delta $ for $n$ running from $%
-7 $ to $7$.

\begin{figure}
\includegraphics[width=1\textwidth]{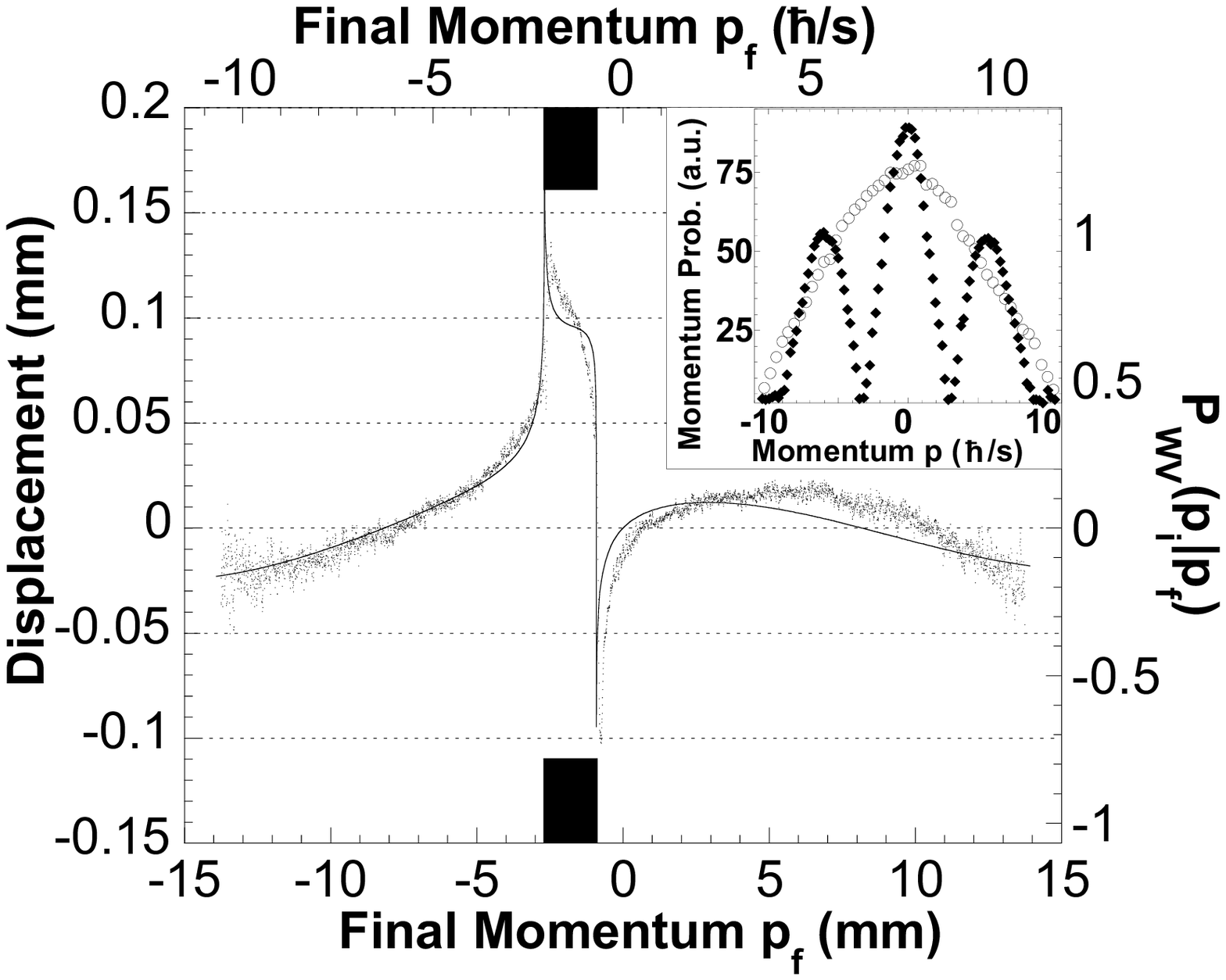}
\caption{ Weak-valued probability $P_{\mathrm{wv}}(p_{i}|p_{f})$, found from 
the $y$-displacement (dots) of the intensity distribution at each $p_f = x_f \cdot (mc/f)$. 
 The thin solid curve
is calculated from Eq.~(\ref{wv prob sum}) following the theory of Ref. %
\cite{Gar04}, using only the independently measured parameters, $w, s$ and $%
\Delta$. The solid black rectangles indicate the weak measurement window, 
$p_{i}\pm \Delta /2$. \red Here $p_{i}=-1.8\mathrm{mm}\cdot (mc/f)$, 
 which is on the side of the central fringe. This gives rise to an  
asymmetric $P_{\mathrm{wv}}(p_{i}|p_{f})$, as explained in the text.
   \blk The inset shows the measured intensity 
for each $p_f$ integrated over $y$, with (solid
diamonds) and without (empty circles) the WWM. The intensity outside
the range shown was below the sensitivity of the CCD.}
\end{figure}

The inset of Fig.~2 shows the momentum distribution of the photons at the
CCD, with and without the WWM, giving $P(p_{f})$ and $P(p_{i})$
respectively. To find $P_{\mathrm{wv}}(p_{i}|p_{f})$,  we measure
 for each $x_{f}$  the average displacement $d$ {in} the $y$-direction of the intensity
distribution while the glass sliver is at $x_{i}$, then divide by $D$.  The example in Fig.~2, for 
$p_{i}=-1.8$mm$\cdot (mc/f)$, shows the typical features of 
$P_{\mathrm{wv}}(p_{i}|p_{f})$. The dominant positive feature of the distribution 
coincides with the window
 $|p_f-p_i|\leq  \Delta/2$. This reflects the fact that half the photons 
suffer no momentum disturbance (see the quantum eraser discussion later). 
The WVP is also positive when $p_f$ is near the minima of the initial interference 
pattern (see inset), as required to ``fill in'' these minima. Similarly, the WVP is 
{\em negative} when $p_f$ is near the maxima of the pattern.
 This negativity proves the existence of a nonclassical momentum disturbance. 
The asymmetry in the curve is because here $p_i$ was chosen to lie on the side of a fringe. 
 Note that diffraction effects due to the non-zero strength
of our weak measurement leads to smoothing of the experimental 
curve in comparison to the theory.

 We sum the conditional
probabilities for all fifteen $p_{i}$ according to Eq.~(\ref{wv prob sum}) to
obtain the unconditional WVP of a momentum transfer $P_{\mathrm{wv}}(q),$ 
plotted in Fig. 3 along with a theoretical curve.  The  agreement between 
 the two is as good as  we expect given the discrepancies in individual data sets 
 exemplified in Fig.~2. 
Our data show that even with
the WWM of the type of Scully \emph{et al.}, $P_{\mathrm{wv}}(q)$ is nonzero outside
the range $\left[ -\hbar /s,\hbar /s\right]$. This supports the 
stance of Storey {\em et al.}  based on their theorem.

 Nonetheless, theory predicts that $P_{\mathrm{wv}}(q)$ has zero variance %
\cite{Wis03}, consistent with the stance of Scully \emph{et al.} 
Unfortunately, as explained in Sec.~3, we cannot obtain the theoretical value 
 of zero because it is practically impossible to obtain data of sufficient quality over 
a sufficient range of momenta to evaluate the required regularized integral \cite{Gar04}. 
 Instead we calculate the integral with sharp cut-offs at $\pm q_{\max}$ (see 
the inset of Fig.~3). The experimental values agree qualitatively with the theoretical curve, which diverges as a function of $q_{\mathrm{max}}$. As explained in Sec.~3, it is 
the oscillations between positive and negative values that ensures 
that the theoretical prediction for the regularized integral is zero. 
The fact that the variance changes sign as a function of $q_{\rm max}$ 
demonstrates that the WWM of the type of Scully \ea\ does not give 
random momentum kicks, and is consistent with the 
weak-valued momentum-transfer variance being zero.

\begin{figure}
\includegraphics[width=1\textwidth]{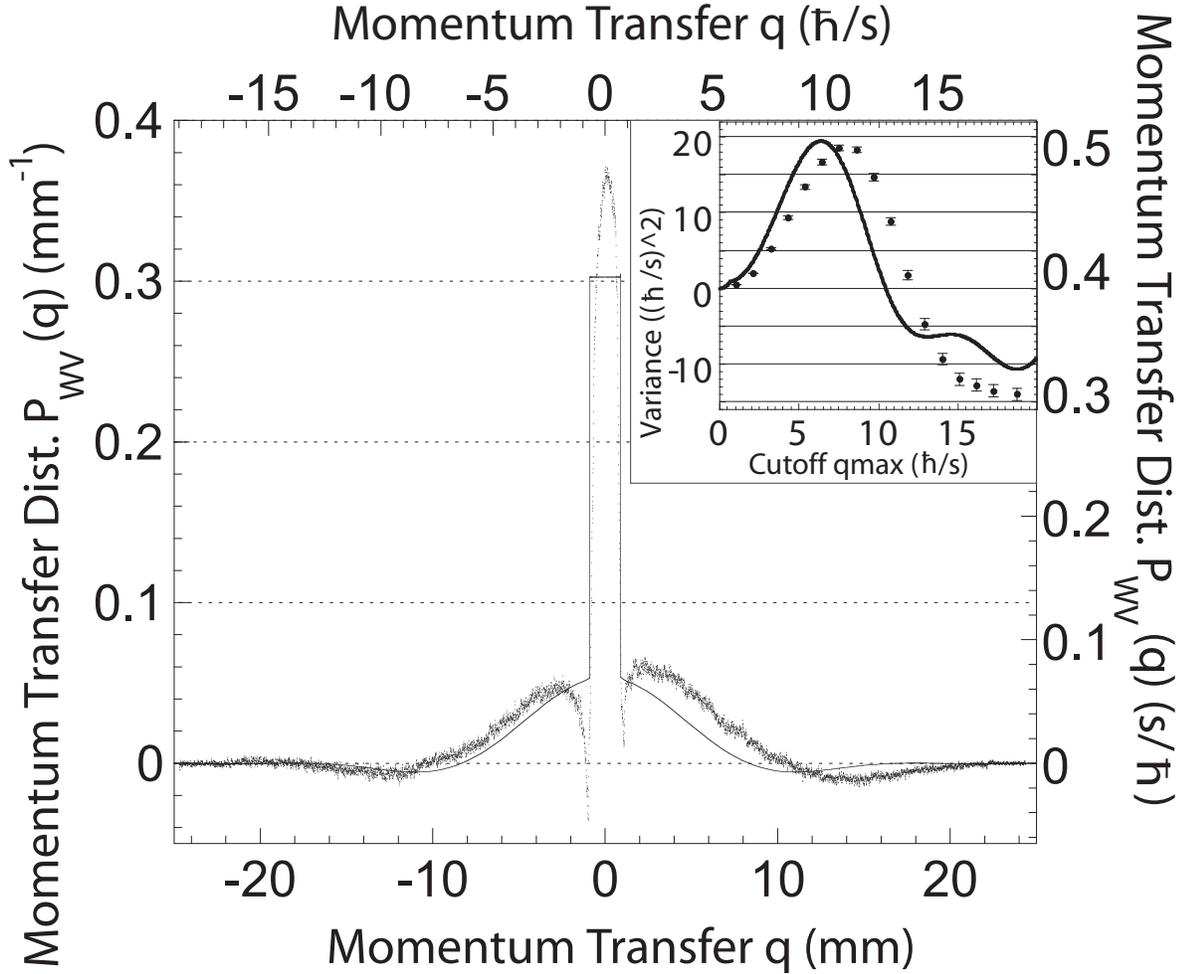}
\caption{ The weak-valued distribution for the momentum transfer %
$P_{\mathrm{wv}}(q)$ calculated from Eq.~(\ref{prob sum}). 
The dots are experimental data and the thin solid
line is theory as in Fig.~2. The inset is the variance integral (experiment as dots,
theory as curve)  over the range $[-q_{\mathrm{max}},q_{\mathrm{max}}]$ 
as a function of $q_{\mathrm{max}}$.}
\end{figure}

Scully \emph{et al.}~\cite{ScuEngWal91} also considered the retrieval of interference in their scheme through the use of a
quantum eraser \cite{ScuDru82}. That is, interference is seen in the subsets of particles selected according to the results of projecting the apparatus in a basis conjugate
to the one that carries the WWM result.  For a WWM with classical momentum transfer,
the different subsets give identical interference patterns apart from being shifted in the $x$-direction by varying amounts \cite{WooZur79}. By contrast, for a WWM 
such as that of Scully \emph{et al.}, the different interference patterns  all have the same {\em envelope}, but with different {\em phases} \cite{Wis97a}.

Our WWM is performed in the horizontal/vertical basis of 
the photon polarization, so we implement a quantum eraser using a polarizer in
the $\pm 45^\circ$ basis. The $45^{\circ }$ photons 
form the usual double-slit interference pattern, whereas the $-45^{\circ }$ photons 
form the antiphase pattern. In Fig.~4 we
plot $P_{\mathrm{wv}}(p_{i}|p_{f})$\ with $p_{i}=-1.8$mm$\cdot (mc/f)$\ for
both polarizer settings, along with the measured interference patterns.
The $45^\circ$ photon data show that,
to a good approximation,  $P_{\mathrm{wv}}(p_{i}|p_{f})=1$ if $|p_{i}-p_{f}|<\Delta /2$ and $0$ otherwise, indicating no momentum
transfer. On the other hand, for the $-45^\circ$ photons, $P_{\mathrm{wv}}(p_{i}|p_{f})$
is substantial even for $p_{f}$ outside the range $p_{i}\pm \Delta /2$. These 
results are found for all values of $p_i$, demonstrating that the momentum transfer
only appears in the photons making the antifringes. This shows 
an intimate connection between the nonclassical momentum transfer and the phase 
between the slits induced by the quantum eraser.

\begin{figure}
\includegraphics[width=0.5\textwidth,angle=-90]{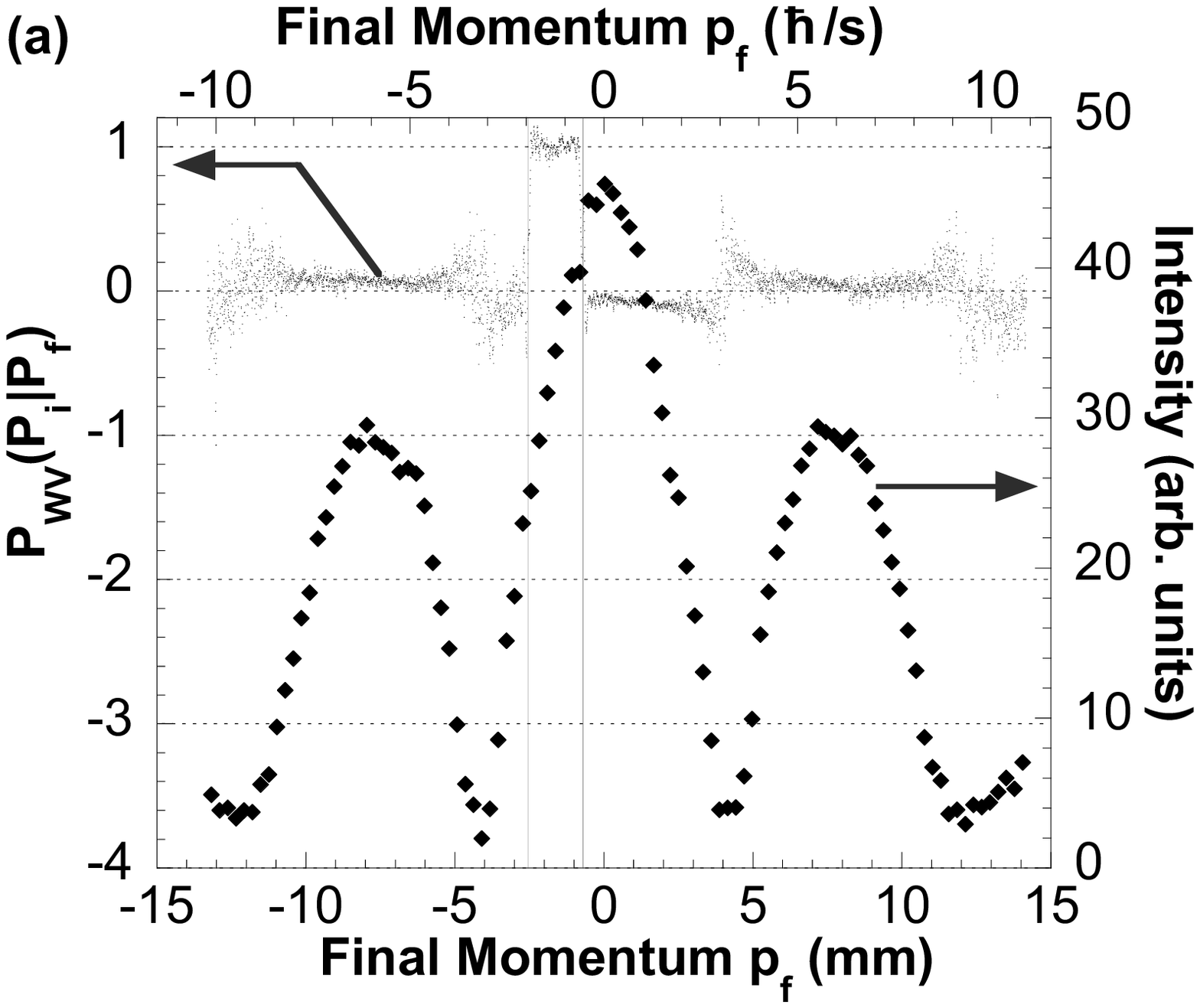} \hspace{-0.25\textwidth}
\includegraphics[width=0.5\textwidth,angle=-90]{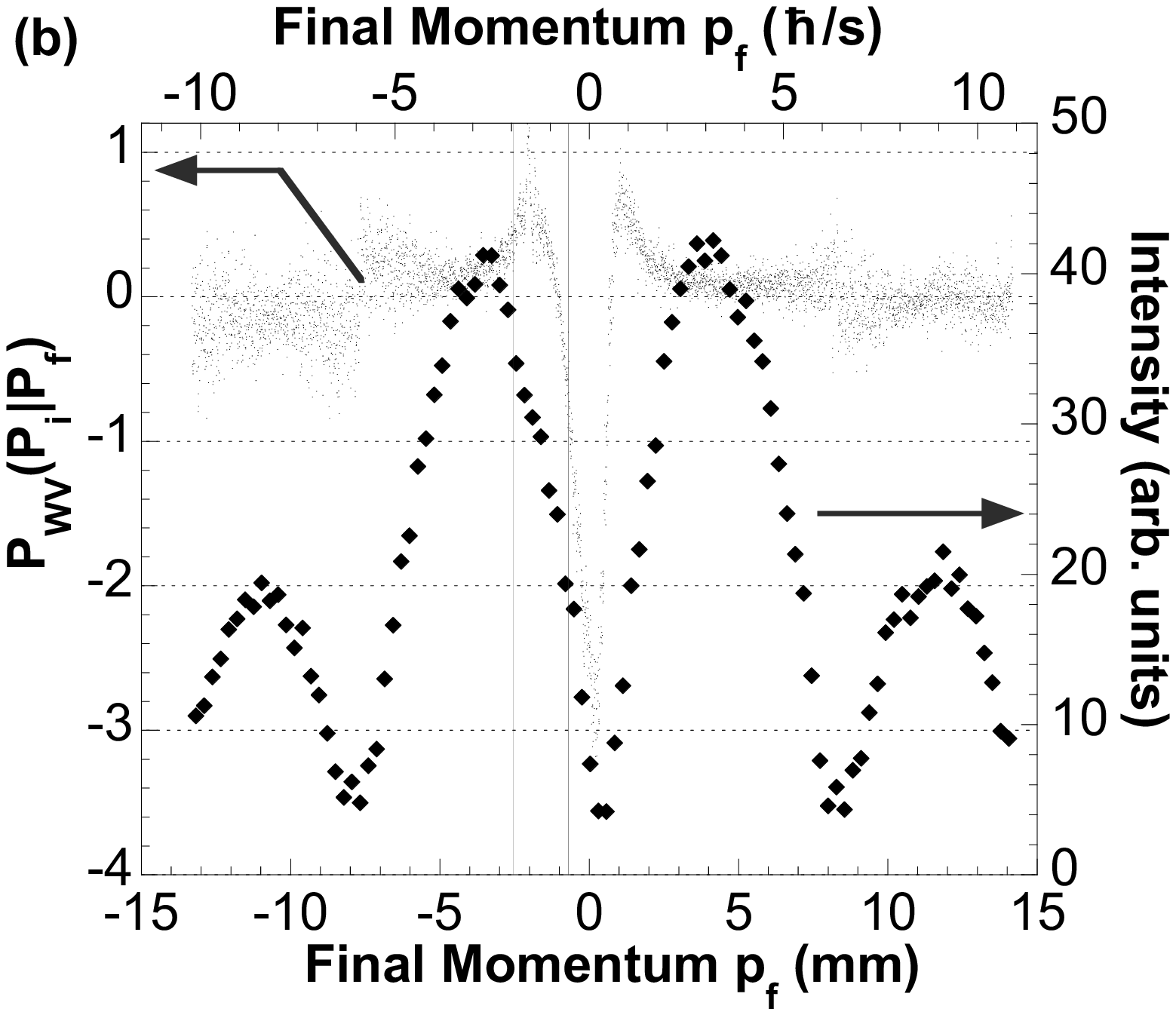}  
\caption{ The WVP $P_{\rm wv}(p_i|p_f)$ for $p_i = -1.8$mm$\cdot (mc/f)$ with a quantum eraser consisting of: a) a $45^\circ$ polarizer; and b) a $-45^\circ$ polarizer, placed after the which-way measurement. The dots indicate the $y$-displacement of the intensity distribution at each $p_f=x_f\cdot (mc/f)$ position on the CCD.   The diamonds indicate the intensity at each $p_f$. The vertical lines indicate $p_i \pm \Delta/2$, the region of the weak measurement.}
\end{figure}

\section{Conclusion}

To conclude, we implemented a WWM of the type Scully \emph{et al.} considered,
and, using the technique of weak measurement, directly observed a distribution for the 
resultant momentum transferred. This distribution spreads well beyond $\pm \hbar/s$, 
in agreement with Storey \emph{et al.}'s claim
that complementarity is a consequence of Heisenberg's 
uncertainty principle (i.e.~the measurement--disturbance relation). 
However, the observed distribution also supports 
Scully \emph{et al.}'s claim of no momentum transfer since its variance 
 is consistent with zero. These seemingly
contradictory observations are compatible only because the weak-valued
 distribution we measure takes negative values, showing the
usefulness of the weak measurement technique in illuminating quantum
processes.

\section*{Acknowledgments} 

This work was supported by the
 ARC, NSERC and PREA. 


\begin{thebibliography}{99}
\bibitem{fn1} In saying that a distribution has width of at least $\sigma$ we
mean it is nonzero somewhere outside the interval $[-\sigma,\sigma]$.
\end{thebibliography}
\end{document}